\documentclass[12pt,graphicx,subfigure,axodraw]{article}
\usepackage{amsfonts}
\usepackage{amssymb}
\usepackage[usenames,dvipsnames]{color}
\usepackage{graphicx}
\usepackage{subfigure}
\usepackage{float}
\def\rmuu{\gamma^{\mu}}
\def\rmud{\gamma_{\mu}}
\def\PL{{1-\gamma_5\over 2}}
\def\PR{{1+\gamma_5\over 2}}
\def\sinW2{\sin^2\theta_W}
\def\AEM{\alpha_{EM}}
\def\mul{M_{\tilde{u} L}^2}
\def\mur{M_{\tilde{u} R}^2}
\def\mdl{M_{\tilde{d} L}^2}
\def\mdr{M_{\tilde{d} R}^2}
\def\mz2{M_{z}^2}
\def\c2b{\cos 2\beta}
\def\au{A_u}
\def\ad{A_d}
\def\cob{\cot \beta}
\def\v#1{v_#1}
\def\tb{\tan\beta}
\def\epem{$e^+e^-$}
\def\KK{$K^0$-$\overline{K^0}$}
\def\wi{\omega_i}
\def\xj{\chi_j}
\def\Wmu{W_\mu}
\def\Wnu{W_\nu}
\def\m#1{{\tilde m}_#1}
\def\mH{m_H}
\def\mw#1{{\tilde m}_{\omega #1}}
\def\mx#1{{\tilde m}_{\chi^{0}_#1}}
\def\mc#1{{\tilde m}_{\chi^{+}_#1}}
\def\mwi{{\tilde m}_{\omega i}}
\def\mxi{{\tilde m}_{\chi^{0}_i}}
\def\mci{{\tilde m}_{\chi^{+}_i}}

\def\ch{{\tilde\chi^{+}_1}}
\def\c2{{\tilde\chi^{+}_2}}

\def\tt{{\tilde\theta}}

\def\tp{{\tilde\phi}}

\def\mz{M_z}
\def\sw{\sin\theta_W}
\def\cw{\cos\theta_W}
\def\cb{\cos\beta}
\def\sb{\sin\beta}
\def\rwi{r_{\omega i}}
\def\rxj{r_{\chi j}}
\def\rfp{r_f'}
\def\Kik{K_{ik}}
\def\Fq2{F_{2}(q^2)}
\def\f{\({\cal F}\)}
\def\d1{{\f(\tilde c;\tilde s;\tilde W)+ \f(\tilde c;\tilde \mu;\tilde W)}}
\def\tw{\tan\theta_W}
\def\sec2w{sec^2\theta_W}
\begin{document}
\baselineskip 18pt
\def\today{\ifcase\month\or
 January\or February\or March\or April\or May\or June\or
 July\or August\or September\or October\or November\or December\fi
 \space\number\day, \number\year}
\def\thebibliography#1{\section*{References\markboth
 {References}{References}}\list
 {[\arabic{enumi}]}{\settowidth\labelwidth{[#1]}
 \leftmargin\labelwidth
 \advance\leftmargin\labelsep
 \usecounter{enumi}}
 \def\newblock{\hskip .11em plus .33em minus .07em}
 \sloppy
 \sfcode`\.=1000\relax}
\let\endthebibliography=\endlist
\def\lsim{\ ^<\llap{$_\sim$}\ }
\def\gsim{\ ^>\llap{$_\sim$}\ }
\def\r2{\sqrt 2}
\def\beq{\begin{equation}}
\def\eeq{\end{equation}}
\def\beqn{\begin{eqnarray}}
\def\eeqn{\end{eqnarray}}
\def\rmuu{\gamma^{\mu}}
\def\rmud{\gamma_{\mu}}
\def\PL{{1-\gamma_5\over 2}}
\def\PR{{1+\gamma_5\over 2}}
\def\sinW2{\sin^2\theta_W}
\def\AEM{\alpha_{EM}}
\def\mul{M_{\tilde{u} L}^2}
\def\mur{M_{\tilde{u} R}^2}
\def\mdl{M_{\tilde{d} L}^2}
\def\mdr{M_{\tilde{d} R}^2}
\def\mz2{M_{z}^2}
\def\c2b{\cos 2\beta}
\def\au{A_u}
\def\ad{A_d}
\def\cob{\cot \beta}
\def\v#1{v_#1}
\def\tb{\tan\beta}
\def\epem{$e^+e^-$}
\def\KK{$K^0$-$\bar{K^0}$}
\def\wi{\omega_i}
\def\xj{\chi_j}
\def\Wmu{W_\mu}
\def\Wnu{W_\nu}
\def\m#1{{\tilde m}_#1}
\def\mH{m_H}
\def\mw#1{{\tilde m}_{\omega #1}}
\def\mx#1{{\tilde m}_{\chi^{0}_#1}}
\def\mc#1{{\tilde m}_{\chi^{+}_#1}}
\def\mwi{{\tilde m}_{\omega i}}
\def\mxi{{\tilde m}_{\chi^{0}_i}}
\def\mci{{\tilde m}_{\chi^{+}_i}}
\def\mz{M_z}
\def\sw{\sin\theta_W}
\def\cw{\cos\theta_W}
\def\cb{\cos\beta}
\def\sb{\sin\beta}
\def\rwi{r_{\omega i}}
\def\rxj{r_{\chi j}}
\def\rfp{r_f'}
\def\Kik{K_{ik}}
\def\Fq2{F_{2}(q^2)}
\def\f{\({\cal F}\)}
\def\d1{{\f(\tilde c;\tilde s;\tilde W)+ \f(\tilde c;\tilde \mu;\tilde W)}}
\def\tw{\tan\theta_W}
\def\sec2w{sec^2\theta_W}
\def\ch{{\tilde\chi^{+}_1}}
\def\c2{{\tilde\chi^{+}_2}}

\def\tt{{\tilde\theta}}

\def\tp{{\tilde\phi}}

\def\mz{M_z}
\def\sw{\sin\theta_W}
\def\cw{\cos\theta_W}
\def\cb{\cos\beta}
\def\sb{\sin\beta}
\def\rwi{r_{\omega i}}
\def\rxj{r_{\chi j}}
\def\rfp{r_f'}
\def\Kik{K_{ik}}
\def\Fq2{F_{2}(q^2)}
\def\f{\({\cal F}\)}
\def\d1{{\f(\tilde c;\tilde s;\tilde W)+ \f(\tilde c;\tilde \mu;\tilde W)}}

\def\b{$\cal{B}(\tau\to\mu \gamma)$~}

\def\tw{\tan\theta_W}
\def\sec2w{sec^2\theta_W}
\newcommand{\pn}[1]{{\color{red}{#1}}}

\begin{titlepage}

~~~~~~~~~~~~~~~~~~~~~~~~~~~~~~~~~~~~~~~~~~~~~~~~~~~~~~~~~~~~~~~~~~~~~~~~~~~~{Preprint No: NSF-KITP-13-080.}
\begin{center}
{\large {\bf
Radiative Decays of  Cosmic Background Neutrinos  in Extensions of MSSM with a Vector Like Lepton Generation}}\\
\vskip 0.5 true cm
Amin Aboubrahim$^{b}$\footnote{Email: amin.b@bau.edu.lb}, Tarek Ibrahim$^{a,b}$\footnote{Email: tarek-ibrahim@alex-sci.edu.eg}
  and Pran Nath$^{c,d}$\footnote{Emal: nath@neu.edu}
\vskip 0.5 true cm
\end{center}

\noindent
{a. Department of  Physics, Faculty of Science,
University of Alexandria,}\\
{ Alexandria 21511, Egypt\footnote{Permanent address.} }\\
{b. Department of Physics, Faculty of Sciences, Beirut Arab University,
Beirut 11 - 5020, Lebanon\footnote{Current address.}} \\
{c. Department of Physics, Northeastern University,
Boston, MA 02115-5000, USA\footnote{Permanent address}} \\
{d. KITP, University of California, Santa Barbara, CA 93106-4030}

\vskip 0.5 true cm

\centerline{\bf Abstract}
An analysis of radiative decays of the neutrinos $\nu_j\to \nu_l \gamma$ is discussed in MSSM extensions with a  vector like lepton generation. Specifically we compute neutrino decays arising from the exchange of charginos and charged sleptons where the photon is emitted by the charged particle in the loop. It is shown that while the lifetime of the neutrino decay in the Standard Model is $\sim 10^{43}$ yrs for a neutrino mass of 50 meV, the current lower limit from experiment from the analysis of the Cosmic Infrared Background is $\sim 10^{12}$ yrs and thus beyond the reach of experiment  in the foreseeable future. However, in the extensions with a vector like lepton generation the lifetime for the decays  can be as low as $\sim 10^{12}- 10^{14}$ yrs and thus within reach of  future improved
experiments. The effect of CP phases on the neutrino lifetime is also analyzed. It is shown that while both the magnetic and the electric transition dipole moments contribute to the neutrino lifetime, often the electric dipole moment dominates even for moderate size CP phases.  \\

\noindent
{\scriptsize
}

\medskip

\end{titlepage}

\section{Introduction}
It is well known  that a neutrino can decay radiatively to neutrinos with lower masses. Thus for the neutrino mass eigenstates $\nu_1$,  $\nu_2$,  $\nu_3$, with $m_{\nu_{3}}> m_{\nu_2}  > m_{\nu_1}$ one can have radiative decays so that $\nu_3 \to \nu_1 \gamma, \nu_2\gamma$.
In the Standard Model this process can proceed by the exchange of a charged lepton and a W boson so that $ \nu_3\to l^- W^+(loop)  \to \nu_{1, 2}\gamma$.
However, the lifetime for the neutrino decay in the Standard Model is rather large~\cite{Pal:1981rm}, i.e.,
\beqn
 \tau_{\nu_3}^{SM}  \sim 10^{43} ~~{\rm yrs},
 \label{1}
  \eeqn
for a $\nu_3$ with mass 50 meV.  Now the current lower limit based on data from galaxy surveys with infrared satellites AKARI~\cite{Matsuura:2010rb}, Spitzer~\cite{Dole:2006de} and Hershel~\cite{Berta:2010rc}  as well as the high precision
cosmic microwave background (CMB) data collected by the Far Infrared Absolute Spectrometer (FIRAS) on board the
Cosmic Background explorer (COBE)~\cite{Mirizzi:2007jd}
 for the study of radiative decays of the cosmic neutrinos\cite{Kim:2011ye} using the Cosmic Infrared Background (CIB) gives \cite{Kim:2011ye}
 \beqn
  \tau_{\nu_3}^{exp}  \geq  10^{12} ~~{\rm yrs}
  \label{2}
  \eeqn
This lower limit is below the Standard Model prediction of Eq.(\ref{1}) by over 30 orders of magnitude
and thus the study of cosmic neutrinos using the Cosmic Infrared Background  is unlikely to be fruitful in
testing the radiative decays of the neutrinos in the Standard Model. However, much lower lifetimes for the
neutrino decays  can be achieved when one goes beyond the Standard Model.
For example, radiative decays of the neutrinos have been discussed in
extensions of the standard model with a heavy mirror generation~\cite{Maalampi:1988vs}.
Using their result one finds a neutrino lifetime $\sim 10^{20}$ yrs which while much smaller than the
one given by the Standard Model is still eight orders of magnitude above
the current level of sensitivity.
Similarly  in the left-right symmetric models,  calculations show that one can lower the lifetime for the decay of the neutrino significantly so that~\cite{Kim:2011ye}  $\tau_{\nu_{3}}^{LR}  \sim 1.5 \times 10^{17} ~~{\rm yrs}$. The experimental measurement using radiative decays provides a way to measure the absolute mass of the neutrino. Thus consider the
decay $\nu_j\to \nu_l\gamma$. In the rest frame of the decay of $\nu_j$ the photon energy is given by
 $E_{\gamma} = (m_j^2- m_l^2)/ (2 m_j)$. Since neutrino oscillations provide us with the neutrino mass difference $m_j^2-m_l^2$, a measurement of the photon energy allows a determination of  $m_j$. Thus the study of  Cosmic Infrared Background provides us  with an alternative way to fix the absolute value of the neutrino mass  aside from the neutrino less double beta decay.\\

In this work we will discuss a new class of models where the neutrino lifetimes as low as close to the current
experimental lower limits can be obtained which makes the study of the lifetimes of the cosmic neutrinos using
CIB interesting. Specifically we consider neutrino decay
via a light vector like generation. Light vector like generations have been discussed in
a variety of works recently.
Specifically these include the neutrino magnetic moments~\cite{Ibrahim:2008gg}, contribution to EDMs of
leptons~\cite{Ibrahim:2010va} and quarks EDMs~\cite{Ibrahim:2011im,Ibrahim:2010hv},
contribution to radiative decay of charged leptons~\cite{Ibrahim:2012ds} and to variety of
other phenomena~\cite{Ibrahim:2009uv,Babu:2008ge,Liu:2009cc,Martin:2009bg,Graham:2009gy}.
Like the flavor changing radiative decay of the charged leptons (for a review see~\cite{Hewett:2012ns} )
the radiative decays of the neutrinos provide a window to new physics.  With the inclusion of the vector generation
we also expect the radiative decays of the neutrinos could be significantly larger than in the Standard Model.
   The reason for this expectation is the following:
   In the analysis of the decay $\tau\to \mu \gamma$ it is found~\cite{Ibrahim:2012ds} that
   the decay for this process is much larger in models with vector like multiplets than
   in conventional models.  We expect that a similar phenomenon will occur in the
   analysis of the radiative decay of the neutrinos.  This is so because the diagrams that enter
in the neutrino radiative decay are very similar to the diagrams that enter in the
analysis of the radiative decay of the $\tau$.
Thus we expect that the analysis would yield a decay lifetime which would be orders of
magnitude closer to the current experimental limits than the result from the Standard Model.
In the analysis we will impose the most recent constraints from the Planck satellite
experiment~\cite{Ade:2013ktc}, i.e., that\footnote{The recent data from the  Planck
experiment~\cite{Ade:2013ktc}, gives two upper limits on the sum of the neutrino masses, i.e.,
0.66 eV and 0.85 eV (both at 95\% CL), where the latter limit includes the lensing likelihood.}
$\sum_{i} m_{\nu_i} < 0.85 {\rm ~eV}   ~(95\% ~CL)$ as well as the neutrino oscillation
constraints~\cite{Schwetz:2008er} on the mass differences $\Delta m^2_{31}\equiv m_3^2-m_1^2= 2.4^{+0.12}_{-0.11} \times 10^{-3}$eV$^2$, and
$\Delta m_{21}^2\equiv m_2^2- m_1^2= 7.65^{+0.23}_{-0.20} \times 10^{-5}$eV$^2$. \\

  We  note in passing that the radiative decays of the cosmic neutrinos in a supersymmetric framework  was discussed
in early work in~\cite{Gabbiani:1990uc}. However,   in their work the radiative decay of neutrinos with testable lifetimes
 make   flavor changing processes in the charged lepton sector
 exceed the experimental limits. Thus these authors had to consider broken R parity models to circumvent these constraints.
In our work there are no problems of this sort in the analysis presented here.  Indeed the flavor changing neutral
currents in the charged sector were already discussed in this class of models in ~\cite{Ibrahim:2012ds}
and the results are consistent with current limits with the possibility of detection of such processes in
 improved experiment.
 The reason why the flavor changing neutral current  processes in the charged sector do not constrain
 the radiative decays of the neutrinos is because
  while  the couplings $f_4, f_4', f_4''$ in Eq.(6) enter the charged lepton sector, they do not enter the neutrino sector.
  Further,
while the couplings $f_5, f_5', f_5''$ enter the neutrino sector they do not enter the charged lepton sector. This allows one to
suppress the neutral current processes in the charged lepton sector without  a problem.
In a similar fashion  the muon  g-2 experiment does not put any constraint on the current analysis. This is so because the contribution of the vector-like multiplet to $g_{\mu} -2$ would arise from couplings $f_4, f_4', f_4''$ which as already
indicated above do not enter
in the radiative decays of the neutrinos and  these couplings can be adjusted
so that the contribution of the vector like multiplet to $g_{\mu}-2$ is  consistent
 with the current $g_\mu-2$ limits. We have not done an explicit analysis of it here since these couplings do not enter in the radiative decays of the neutrinos and hence are not relevant for the analysis of this paper.
  \\

\section{Extension of MSSM with a vector multiplet}

Vector like multiplets arise in a variety of unified models~\cite{vectorlike} some of which could be low lying. Here we simply assume
the existence of one low lying leptonic vector multiplet which is anomaly free  in addition to the MSSM spectrum.  Before proceeding further
it is useful to record the  quantum numbers of the leptonic matter content of this extended MSSM spectrum under $SU(3)_C\times SU(2)_L\times U(1)_Y$.
Thus  under $SU(3)_C\times SU(2)_L\times U(1)_Y$ the leptons of the
 three generations transform as follows
\beqn
\psi_{iL}\equiv
 \left(\matrix{ \nu_{i L}\cr
 ~{l}_{iL}}\right)
\sim(1,2,- \frac{1}{2}), l^c_{iL}\sim (1,1,1), ~\nu^c_{i L}\sim (1,1,0), ~i=1,2,3
\label{2}
\eeqn
where the last entry on the right hand side of each $\sim$ is the value of the hypercharge
 $Y$ defined so that $Q=T_3+ Y$.  These leptons have $V-A$ interactions.
We can now add a vector like multiplet where we have a fourth family of leptons with $V-A$ interactions
whose transformations can be gotten from Eq.(\ref{2}) by letting i run from 1-4.
A vector like lepton multiplet also has  mirrors and so we consider these mirror
leptons which have $V+A$ interactions. Their quantum numbers are
as follows
\beqn
\chi^c\equiv
 \left(\matrix{ E_{ L}^c\cr
 N_L^c}\right)
\sim(1,2,\frac{1}{2}), E_{ L}\sim (1,1,-1), N_L\sim (1,1,0).
\label{3}
\eeqn
 The MSSM Higgs  doublets as usual have the quantum numbers
\beqn
H_1\equiv
 \left(\matrix{ H_1^1\cr
 H_1^2}\right)
\sim(1,2,-\frac{1}{2}), ~H_2\equiv
 \left(\matrix{ H_2^1\cr
 H_2^2}\right)
\sim(1,2,\frac{1}{2}).
\label{4}
\eeqn

As mentioned already we assume that the vector  multiplet  escapes acquiring mass at the GUT scale and
remains light down to the electroweak scale.
As in the analysis  of Ref.\cite{Ibrahim:2010va}
interesting new physics arises when we consider the
mixing of the second and third generations of leptons with the mirrors of the vector like multiplet.
Actually  we will extend our model to include the mixing of the first generation as well, for the  computation of the decay $\nu_3 \to \nu_{2, 1} \gamma$. Thus the  superpotential of the model may be written
in the form
\beqn
W= -\mu \epsilon_{ij} \hat H_1^i \hat H_2^j+\epsilon_{ij}  [f_{1}  \hat H_1^{i} \hat \psi_L ^{j}\hat \tau^c_L
 +f_{1}'  \hat H_2^{j} \hat \psi_L ^{i} \hat \nu^c_{\tau L}
+f_{2}  \hat H_1^{i} \hat \chi^c{^{j}}\hat N_{L}
 +f_{2}'  H_2^{j} \hat \chi^c{^{i}} \hat E_{ L}\nonumber\\
+ h_{1}  H_1^{i} \hat\psi_{\mu L} ^{j}\hat\mu^c_L
 +h_{1}'  H_2^{j} \hat\psi_{\mu L} ^{i} \hat\nu^c_{\mu L}
+ h_{2}  H_1^{i} \hat\psi_{e L} ^{j}\hat e^c_L
 +h_{2}'  H_2^{j} \hat\psi_{e L} ^{i} \hat\nu^c_{e L}]\nonumber\\
+ f_{3} \epsilon_{ij}  \hat\chi^c{^{i}}\hat\psi_L^{j}
 + f_{3}' \epsilon_{ij}  \hat\chi^c{^{i}}\hat\psi_{\mu L}^{j}\nonumber\\
 + f_{4} \hat\tau^c_L \hat E_{ L}  +  f_{5} \hat\nu^c_{\tau L} \hat N_{L}
 + f_{4}' \hat\mu^c_L \hat E_{ L}  +  f_{5}' \hat\nu^c_{\mu L} \hat N_{L}\nonumber\\
+ f_{3}'' \epsilon_{ij}  \hat\chi^c{^{i}}\hat\psi_{e L}^{j}
 + f_{4}'' \hat e^c_L \hat E_{ L}  +  f_{5}'' \hat\nu^c_{e L} \hat N_{L}
 \label{5}
\eeqn
where $\hat\psi_L$ stands for $\hat\psi_{3L}$, $\hat\psi_{\mu L}$ stands for $\hat\psi_{2L}$
and  $\hat\psi_{e L}$ stands for $\hat\psi_{1L}$.
Here we assume a mixing between the mirror generation and the third lepton generation through
the couplings $f_3$, $f_4$ and $f_5$. We also assume mixing between the mirror generation and the
second lepton generation through the couplings $f_3'$, $f_4'$ and $f_5'$. The same is true for the mixing between the
mirror generation and the first lepton generation through the couplings $f_3''$, $f_4''$ and $f_5''$.
{\it The above nine mass terms are responsible for generating lepton flavor changing process.}
We will focus here on the supersymmetric  sector. Then through the terms $f_3, f_4, f_5, f_3', f_4', f_5', f_3'', f_4'', f_5''$
 one can have a mixing between
the third generation, the second and the first generation leptons
which allows the decay of $\nu_{3}\to \nu_{2, 1}  \gamma$  through loop
corrections that include charginos  and scalar lepton exchanges
with the photon being emitted by the chargino
or by a charged slepton.
The mass terms for the leptons and  mirrors arise from the term
\beq
{\cal{L}}=-\frac{1}{2}\frac{\partial ^2 W}{\partial{A_i}\partial{A_j}}\psi_ i \psi_ j+H.c.
\label{6}
\eeq
where $\psi$ and $A$ stand for generic two-component fermion and scalar fields.
After spontaneous breaking of the electroweak symmetry, ($\langle H_1^1 \rangle=v_1/\sqrt{2} $ and $\langle H_2^2\rangle=v_2/\sqrt{2}$),
we have the following set of mass terms written in the 4-component spinor notation
\beqn
-{\cal L}_m =
 \left(\matrix{\bar \nu_{\tau R} & \bar N_R & \bar \nu_{\mu R}
&\bar \nu_{e R} }\right)
 \left(\matrix{f'_1 v_2/\sqrt{2} & f_5 & 0 & 0 \cr
 -f_3 & f_2 v_1/\sqrt{2} & -f_3' & -f_3'' \cr
0&f_5'&h_1' v_2/\sqrt{2} & 0 \cr
0 & f_5'' & 0 & h_2' v_2/\sqrt{2} }\right)\left(\matrix{ \nu_{\tau L}\cr
 N_L \cr
\nu_{\mu L}\cr
\nu _{e L}}\right)  + H.c.
\label{7}
\eeqn
Here
the mass matrices are not  Hermitian and one needs
to use bi-unitary transformations to diagonalize them. Thus we write the linear transformations
\beqn
 \left(\matrix{ \nu_{\tau_R}\cr
 N_{ R} \cr
\nu_{\mu_R} \cr
\nu_{e_R}}\right)=D^{\nu}_R \left(\matrix{ \psi_{1_R}\cr
 \psi_{2_R}  \cr
\psi_{3_R} \cr
\psi_{4_R}}\right),\nonumber\\
\left(\matrix{ \nu_{\tau_L}\cr
 N_{ L} \cr
\nu_{\mu_L} \cr
\nu_{e_L} }\right)=D^{\nu}_L \left(\matrix{ \psi_{1_L}\cr
 \psi_{2_L} \cr
\psi_{3_L} \cr
\psi_{4_L}}\right),
\label{8}
\eeqn
such that
\beq
D^{\nu \dagger}_R
 \left(\matrix{f'_1 v_2/\sqrt{2} & f_5 & 0 & 0 \cr
 -f_3 & f_2 v_1/\sqrt{2} & -f_3' & -f_3'' \cr
0&f_5'&h_1' v_2/\sqrt{2} & 0 \cr
0 & f_5'' & 0 & h_2' v_2/\sqrt{2} }\right)
 D^{\nu}_L=diag(m_{\psi_1},m_{\psi_2},m_{\psi_3}, m_{\psi_4} ).
\label{put2}
\label{10}
\eeq
{
In Eq.(\ref{10})
$\psi_1, \psi_2, \psi_3, \psi_4$ are the mass eigenstates for the neutrinos,
where in the limit of no mixing
we identify $\psi_1$ as the light tau neutrino, $\psi_2$ as the
heavier mass eigen state,  $\psi_3$ as the muon neutrino and $\psi_4$ as the electron neutrino.
To make contact with the normal neutrino hierarchy we relabel the states so that

\beqn
\nu_1= \psi_4, \nu_2= \psi_3, \nu_3= \psi_1, \nu_4= \psi_2
\eeqn
which we assume has the mass hierarchical pattern

\beqn
m_{\nu_1}< m_{\nu_2} < m_{\nu_3} < m_{\nu_4}
\eeqn
We will carry out the analytical analysis in the $\psi_i$ notation but the numerical analysis
will be carried out in the $\nu_i$ notation to make direct contact with data.
  Next we  consider  the mixing of the charged sleptons and the charged mirror sleptons.
The mass squared  matrix of the slepton - mirror slepton comes from three sources, the F term, the
D term of the potential and the soft susy breaking terms.
Using the  superpotential of Eq.(\ref{5}) the mass terms arising from it
after the breaking of  the electroweak symmetry are given by
the Lagrangian
\beq
{\cal L}= {\cal L}_F +{\cal L}_D
\eeq
where   $ {\cal L}_F$ and $ {\cal L}_D$ are given in the Appendix along with the matrix elements of the slepton mass squared  matrix.

\section{Interactions of charginos, sleptons and neutrinos}

The chargino exchange contribution to the decay of  the tau neutrino  into a muon neutrino (electron neutrino) and a photon arises through
the  loop diagram in Fig.(1).   The relevant part of  the Lagrangian that
 generates this contribution is given by
\beqn
-{\cal{L}}_{\nu-\tilde{\tau}-\chi^+}=
\sum_{j =1}^4\sum_{i=1}^2\sum_{k=1}^8
\bar{\psi}_{j}[C^L_{j i k} P_L+
 C^R_{j i k} P_R]
\tilde{\chi}^+_i \tilde{\tau}_k +H.c.
\label{16}
\eeqn
where
\beqn
C^L_{j i k}=-f'_1  V^*_{i2} D^{\nu *}_{R _{1 j}}\tilde{D}^{\tau}_{1 k}
-f'_2  V^*_{i2} D^{\nu *}_{R _{2 j}}\tilde{D}^{\tau}_{2 k}\nonumber\\
+g  V^*_{i1} D^{\nu *}_{R _{2 j}}\tilde{D}^{\tau}_{4 k}
-h'_1  V^*_{i2} D^{\nu *}_{R _{3 j}}\tilde{D}^{\tau}_{5 k}
-h'_2  V^*_{i2} D^{\nu *}_{R _{4 j}}\tilde{D}^{\tau}_{7 k},
\nonumber\\
C^R_{j i k}=-f_1  U_{i2} D^{\nu *}_{L _{1 j}}\tilde{D}^{\tau}_{3 k}
-h_1  U_{i2} D^{\nu *}_{L _{3 j}}\tilde{D}^{\tau}_{6 k}
+g  U_{i1} D^{\nu *}_{L _{1 j}}\tilde{D}^{\tau}_{1 k}\nonumber\\
+g  U_{i1} D^{\nu *}_{L _{4 j}}\tilde{D}^{\tau}_{7 k}
-h_2  U_{i2} D^{\nu *}_{L _{4 j}}\tilde{D}^{\tau}_{8 k}
-f_2  U_{i2} D^{\nu *}_{L _{2 j}}\tilde{D}^{\tau}_{4 k},
\label{17}
\eeqn
where $\tilde{D}^{\tau}$ is the diagonalizing matrix of the scalar $8\times 8$ mass squared  matrix
for the scalar leptons as defined
in the Appendix.
In Eq.(\ref{17})
 $U$ and $V$ are the matrices  that  diagonalize the chargino mass matrix $M_C$
  so that
\beq
U^* M_C V^{-1}= diag (m_{\tilde{\chi_1}}^+,m_{\tilde{\chi_2}}^+).
\label{19}
\eeq

\begin{figure}[h]
\begin{center}
{\rotatebox{0}{\resizebox*{13cm}{!}{\includegraphics{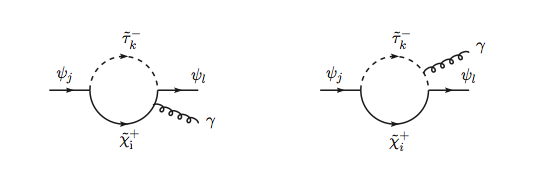}}\hglue5mm}}
\caption{The diagrams that allow decay of the $\psi_j$ into $\psi_l+\gamma$ via supersymmetric loops involving the charginos and the staus where the photon is either emitted by the chargino (left) or by the stau (right) inside the loop.} \label{taumugamma}
\end{center}
\end{figure}

\section{The analysis of $\psi_j \rightarrow \psi_l + \gamma$  decay width }

The decay $\psi_j \rightarrow \psi_l + \gamma$
is induced by one-loop electric and
magnetic  transition dipole moments,  which arise
from the diagrams of Fig.(1). In the dipole moment loop, the incoming $\psi_j$ is replaced by a $\psi_l$.
For an incoming   $\psi_j$ of momentum $p$ and a resulting $\psi_l$ of momentum $p'$, we define the amplitude
\beq
\langle \psi_l (p') | J_{\alpha} | \psi_j (p)\rangle = \bar{u}_{\psi_l} (p') \Gamma_{\alpha} u_{\psi_j} (p)
\label{26}
\eeq
where
\beq
\Gamma_{\alpha} (q) =\frac{F^{j l}_2 (q) i \sigma_{\alpha \beta} q^{\beta}}{m_{\psi_j} +m_{\psi_l}}
+\frac{F^{j l}_3 (q)  \sigma_{\alpha \beta} \gamma_5 q^{\beta}}{m_{\psi_j} +m_{\psi_l}}+.....
\label{27}
\eeq
with $q = p -p'$ and where $m_f$ denotes the mass of the fermion $f$.
The decay width of $\psi_j \rightarrow \psi_l + \gamma$ is given by
\beqn
   \Gamma (\psi_j \rightarrow \psi_l + \gamma) =\frac{m^3_{\psi_j}}{8 \pi  (m_{\psi_j}+m_{\psi_l})^2}
\left(1-\frac{m^2_{\psi_l}}{m^2_{\psi_j}}\right)^3
 \{|F^{j l }_2 (0)|^2
+|F^{j l}_3 (0)|^2 \}
\label{28}
\eeqn
where the form factors $F^{j l}_2$ and $F^{j l}_3$ arise from the left and the
right loops of Fig. (1) as follows
\beqn
F^{j l}_2 (0) = F^{j l}_{2~ left} + F^{j l}_{2~ right} \nonumber\\
F^{j l}_3 (0) = F^{j l}_{3~ left} + F^{j l}_{3~ right}
\label{29}
\eeqn

\noindent
The chargino contribution  $F^{j l}_{2 ~left}$ is given by
\beqn
F^{j l}_{2 ~left}=-\sum_{i=1}^2 \sum_{k=1}^8
\Bigg[\frac{(m_{\psi_j} +m_{\psi_l})}{64 \pi^2 m_{\tilde{\chi_i}^+}}
\{C^L_{l ik} C^{R*}_{j ik} + C^R_{lik} C^{L*}_{jik} \} F_3 \left(\frac{M^2_{\tilde{\tau_k}}}{m^2_{\tilde{\chi_i}^+}}\right)\nonumber \\
+ \frac{m_{\psi_j}(m_{\psi_j} +m_{\psi_l})}{192 \pi^2 m^2_{\tilde{\chi_i}^+}}
\{C^L_{l ik} C^{L*}_{j ik} + C^R_{lik} C^{R*}_{jik}  \} F_4 \left(\frac{M^2_{\tilde{\tau_k}}}{m^2_{\tilde{\chi_i}^+}}\right)\Bigg]
\label{30}
\eeqn
where
\beq
F_3(x)= \frac{1}{(x-1)^3} \{3 x^2-4x +1 -2x^2 \ln x \}
\label{31}
\eeq
and
\beq
F_4(x)= \frac{1}{(x-1)^4} \{2 x^3+3x^2 -6x +1 -6x^2 \ln x \}
\label{32}
\eeq

\noindent
The right contribution  $F^{j l}_{2~right}$ is given by
\beqn
F^{j l}_{2~right}= \sum_{i=1}^2\sum_{k=1}^8
 \Bigg[ \frac{(m_{\psi_j} +m_{\psi_l})}{64 \pi^2 m_{\tilde{\chi_i}^+}}
\{C^L_{l ik} C^{R*}_{j ik} + C^R_{lik} C^{L*}_{jik} \} F_1 \left(\frac{M^2_{\tilde{\tau_k}}}{m^2_{\tilde{\chi_i}^+}}\right)\nonumber\\
+ \frac{m_{\psi_j}(m_{\psi_j} +m_{\psi_l})}{192 \pi^2 m^2_{\tilde{\chi_i}^+}}
\{C^L_{l ik} C^{L*}_{j ik} + C^R_{lik} C^{R*}_{jik}  \} F_2 \left(\frac{M^2_{\tilde{\tau_k}}}{m^2_{\tilde{\chi_i}^+}}\right)\Bigg]
\label{33}
\eeqn
where
\beq
F_1(x)= \frac{1}{(x-1)^3} \{1-x^2  +2x \ln x \}
\label{34}
\eeq
and
\beq
F_2(x)= \frac{1}{(x-1)^4} \{- x^3 +6x^2 -3x -2 -6x \ln x \}
\label{35}
\eeq

\noindent
The left contribution  $F^{j l}_{3~left}$ is given by
\beqn
F^{j l}_{3 ~left}= -\sum_{i=1}^2 \sum_{k=1}^8
 \frac{(m_{\psi_j} +m_{\psi_l})m_{\tilde{\chi_i}^+} }{32 \pi^2 M^2_{\tilde{\tau_k}}}
\{ C^L_{j i k} C^{R*}_{l i k} - C^R_{j i k} C^{L*}_{l i k} \}
F_6\left(\frac{m^2_{\tilde{\chi_i}^+}}{M^2_{\tilde{\tau_k}}}\right)
\label{36}
\eeqn
where
\beq
F_6(x)= \frac{1}{2(x-1)^2} \Bigg\{-x +3 + \frac{2\ln x}{1-x}\Bigg\}
\label{37}
\eeq

The right contribution  $F^{j l}_{3~right}$ is given by
\beqn
F^{j l}_{3 ~right}= \sum_{i=1}^2 \sum_{k=1}^8
 \frac{(m_{\psi_j} +m_{\psi_l})m_{\tilde{\chi_i}^+} }{32 \pi^2 M^2_{\tilde{\tau_k}}}
\{ C^L_{j i k} C^{R*}_{l i k} - C^R_{j i k} C^{L*}_{l i k} \}
F_5\left(\frac{m^2_{\tilde{\chi_i}^+}}{M^2_{\tilde{\tau_k}}}\right)
\label{36}
\eeqn
where
\beq
F_5(x)= \frac{1}{2(x-1)^2} \Bigg\{1+x+ \frac{2 x\ln x}{1-x} \Bigg\}
\label{37}
\eeq

Now for the numerical analysis below we switch from the $\psi_i$ notation to the $\nu_i$ notation.
Here $\nu_1, \nu_2, \nu_3$ are the three neutrino mass eigenstates and we assume the mass hierarchy so
that $\nu_3$ is heavier than $\nu_2$ and $\nu_2$ is heavier than $\nu_1$. For the cosmic neutrinos
we are interested in the decay of the  $\nu_3$  to $\nu_2$ and $\nu_1$.
Thus the  total decay width  of $\nu_3$ is given by  $\Gamma_{total}(\nu_3) =\Gamma(\nu_3 \rightarrow \nu_2 +\gamma) +\Gamma(\nu_3 \rightarrow \nu_1 +\gamma)$. The lifetime of the tau neutrino is calculated from the equation
\beqn
\tau (\nu_3) = \frac{\hbar}{ \Gamma_{total}(\nu_3)}
\label{38}
\eeqn
where the $\Gamma_{total}(\nu_3)$ is in GeV and
$\hbar =2.085\times10^{-32}$ GeV.Year.

\noindent

\section{ Estimates of $\nu_3$  lifetime}

In this section we give  a numerical  estimate of the neutrino lifetime for the heaviest neutrino $\nu_3$ and investigate its dependence on the input parameters.
In the analysis we  ensure that the
 constraint of $\Sigma_i m_{\nu_i} < 0.85$ eV from the Planck  Satellite experiment~\cite{Ade:2013ktc} is satisfied and
that $\Delta m_{31}^2$
and  $\Delta m_{21}^2$
 lie in the $3\sigma$ range of the neutrino oscillation experiment~\cite{Schwetz:2008er},  i.e., in the range of $(2.07-2.75)\times 10^{-3}$ eV$^2$ and  $(7.05-8.34)\times 10^{-5}$ eV$^2$ respectively.
In Table (\ref{table1}), we give a benchmark point where the constraints mentioned above are satisfied.
The form factors and the lifetime of the $\nu_3$ decay are calculated and given in Table (\ref{table1}).
\begin{table}[h]
\begin{center}
\begin{tabular}{|c|c|c|}
  \hline
  & Neutrino Mass Eigenvalues (GeV)  &$m_{\nu_{3}}=5.232137\times 10^{-11}$ \\
       &                           & $m_{\nu_{2}}=8.517946\times 10^{-12}$  \\
    &                          & $m_{\nu_{1}}=1.036377\times 10^{-12}$ \\
     \hline
Process:     & $F_{2~left}^{jl}$ & $(1.4036\times 10^{-20})\exp(-2.73\,i)$ \\ 
    & $F_{2~right}^{jl}$ & $(1.6163\times 10^{-20})\exp(+0.42\,i)$ \\
$\nu_{3}\rightarrow \nu_{2}+\gamma$    & $F_{2}^{jl}(0)$ & $(2.1357\times 10^{-21})\exp(+0.51\,i)$ \\
    & $F_{3~left}^{jl}$ & $(7.6091\times 10^{-18})\exp(+2.42\,i)$ \\ 
    & $F_{3~right}^{jl}$ & $(1.8846\times 10^{-18})\exp(+2.42\,i)$ \\
    & $F_{3}^{jl}(0)$ & $(9.4946\times 10^{-18})\exp(+2.42\,i)$ \\
    & Decay Width & $1.2802755\times 10^{-46}$ GeV  \\
    \hline
Process: & $F_{2~left}^{jl}$ & $(2.9501\times 10^{-21})\exp(+1.57\,i)$ \\ 
   & $F_{2~right}^{jl}$ & $(2.8460\times 10^{-20})\exp(+0.37\,i)$ \\
$\nu_{3}\rightarrow \nu_{1}+\gamma$ & $F_{2}^{jl}(0)$ & $(2.9655\times 10^{-20})\exp(+0.46\,i)$ \\
   & $F_{3~left}^{jl}$ & $(1.0064\times 10^{-18})\exp(-1.77\,i)$ \\ 
   & $F_{3~right}^{jl}$ & $(2.4903\times 10^{-19})\exp(-1.77\,i)$ \\
   & $F_{3}^{jl}(0)$ & $(1.2555\times 10^{-18})\exp(-1.77\,i)$ \\
   & Decay Width & $3.1531459\times 10^{-48}$ GeV \\
   \hline
   & Life time & $1.5899\times 10^{14}$ {Years} \\
  \hline
\end{tabular}
\caption{
Sample  numerical values for the neutrino masses and the calculated form factors and decay widths of the two processes $\nu_{3}\rightarrow \nu_{2}+\gamma$ and $\nu_{3}\rightarrow \nu_{1}+\gamma$. The  lifetime is also given. The analysis corresponds to the parameter set:
 $|m_{2}|=150$, $|\mu|=100$, $|f_{3}|=1.5\times10^{-7}$, $|f_{3}'|=2\times10^{-8}$, $|f_{3}''|=8\times10^{-9}$, $|f_{4}|=|f_{4}'|=|f_{4}''|=50$, $|f_{5}|=8.11\times10^{-2}$, $|f_{5}'|=9.8\times10^{-2}$, $|f_{5}''|=4\times10^{-2}$, $m_{N}=212$, $|A_{0}|=600$, $m_{E}=260$, $m_{0}=300$, $\tan\beta=50$, $\chi_{m_{2}}=1.2$, $\chi_{\mu}=0.8$, $\chi_{3}=0.3$, $\chi_{3}'=0.2$, $\chi_{3}''=0.6$, $\chi_{4}=1.4$, $\chi_{4}'=1.1$, $\chi_{4}''=1.7$, $\chi_{5}=1.7$, $\chi_{5}'=0.5$, $\chi_{5}''=0.7$ and $\chi_{A_{0}}=2.4$. All masses are in GeV and phases in rad.}
 \label{table1}
\end{center}
\end{table}

We now begin by exhibiting the dependence of the $\nu_3$ lifetime on the $SU(2)$ gaugino mass $m_2$. The chargino masses
are sensitive to $m_2$ and increasing $m_2$ implies a larger average chargino mass which  affects the $\nu_3$  decay width and  the
lifetime. This is exhibited in  Fig.~(\ref{m2}) for values of $\tan\beta$ = 30, 40, 50 while the values of the other input parameters are shown in the caption of Fig.~(\ref{m2}). It  is found that both the magnetic and the electric transition dipole moments enter in the analysis. The magnetic transition dipole moment depends on  $F_2^{jl}$ while  the electric transition dipole moment depends on $F_3^{jl}$. Typically the electric transition dipole moment
dominates the decay even for  moderate size CP phases since $F_3^{jl}$ turns out to be much larger than $F_2^{jl}$.\\

\begin{figure}[h]
\begin{center}
{\rotatebox{0}{\resizebox*{12cm}{!}{\includegraphics{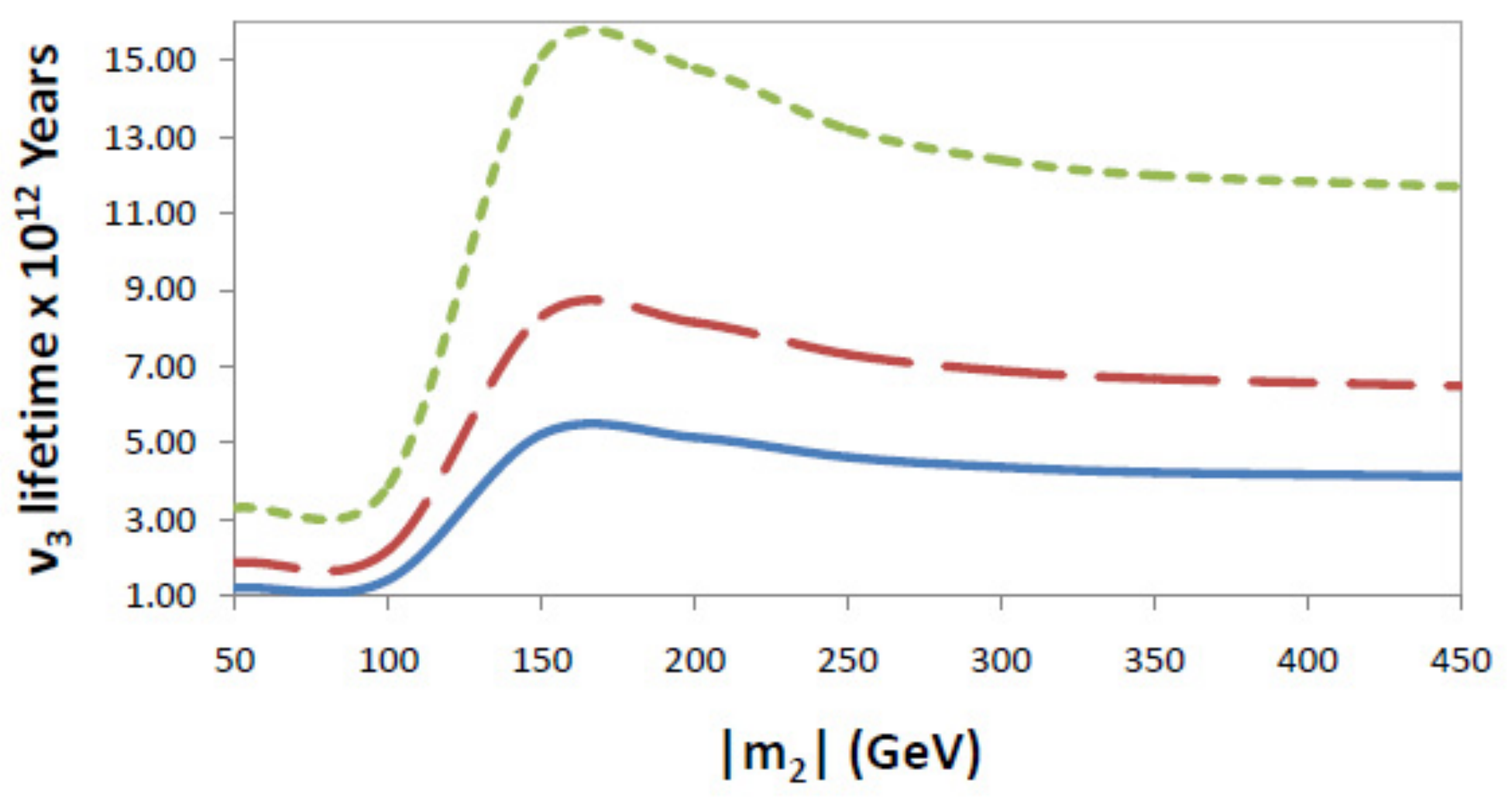}}\hglue5mm}}
\caption{Variation of {$\nu_3$} lifetime versus $|m_{2}|$ for three values of $\tan\beta$. Starting with the upper curve, $\tan\beta$ = 30, 40, 50.  Other parameters have the values
$|\mu|=100$, $|f_{3}|=1.5\times10^{-7}$, $|f_{3}'|=2\times10^{-8}$, $|f_{3}''|=8\times10^{-9}$, $|f_{4}|=|f_{4}'|=|f_{4}''|=35$, $|f_{5}|=1.01\times10^{-1}$, $|f_{5}'|=5.3\times10^{-1}$, $|f_{5}''|=4\times10^{-2}$, $m_{N}=200$, $|A_{0}|=500$, $m_{E}=260$, $m_{0}=300$, $\chi_{m_{2}}=1.2$, $\chi_{\mu}=0.8$, $\chi_{3}=0.3$, $\chi_{3}'=0.2$, $\chi_{3}''=0.6$, $\chi_{4}=1.4$, $\chi_{4}'=1.1$, $\chi_{4}''=1.7$, $\chi_{5}=1.7$, $\chi_{5}'=0.5$, $\chi_{5}''=0.7$ and $\chi_{A_{0}}=0.4$. All masses are in GeV and phases in rad.}
\label{m2}
\end{center}
\end{figure}

In Fig.~(\ref{m0}) we investigate the effect of the variation of $m_{0}$ on $\nu_3$
 lifetime, where  $m_{0}^{2}=\tilde{M}_{\tau L}^{2}=\tilde{M}_{E}^{2}=\tilde{M}_{\tau}^{2}=\tilde{M}_{\chi}^{2}=\tilde{M}_{\mu L}^{2}=\tilde{M}_{\mu}^{2}=\tilde{M}_{eL}^{2}=\tilde{M}_{e}^{2}$ (see Appendix).   Three curves are shown on the figure, corresponding to $\tan\beta$ = 30, 40, 50,
starting from the upper curve ($\tan\beta=30$) and going down. The analysis shows that the lifetime of $\nu_3$ increases as $m_0$ increases. This is as expected
since a larger $m_0$ implies larger sfermion masses that enter in the loop which gives a smaller decay width and a larger lifetime. It is seen that with
values of the input parameters in reasonable ranges the lifetime can be as low as few times $10^{12}$ yrs just within the reach of improved CIB  experiment.\\

\begin{figure}[h]
\begin{center}
{\rotatebox{0}{\resizebox*{12cm}{!}{\includegraphics{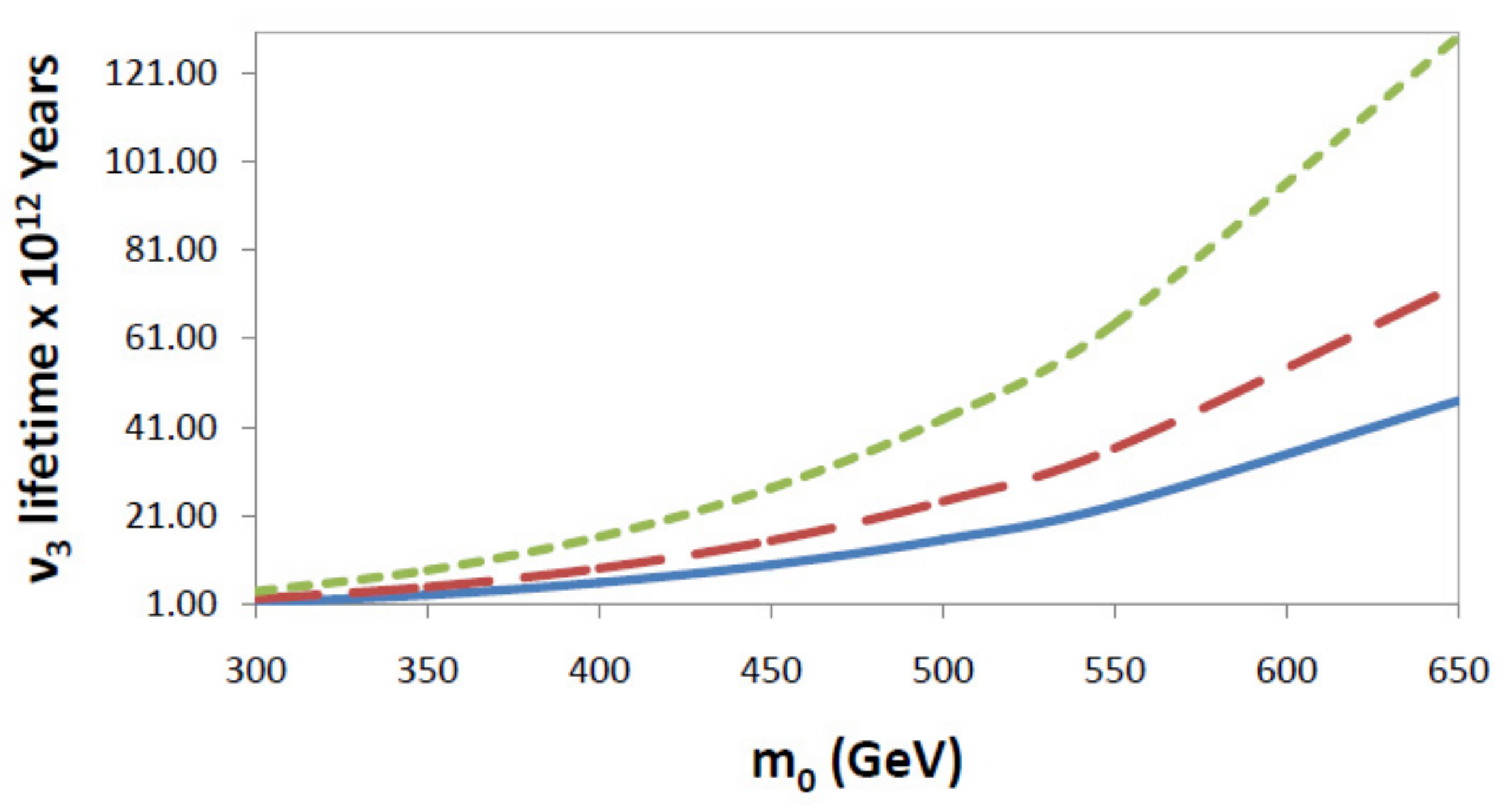}}\hglue5mm}}
\caption{Exhibition of the dependence of {$\nu_3$} lifetime on $m_{0}$ for three values of $\tan\beta$. Starting with the upper curve, $\tan\beta$ = 30, 40, 50. Other parameters have the values
$|\mu|=100$, $|f_{3}|=1.5\times10^{-7}$, $|f_{3}'|=2\times10^{-8}$, $|f_{3}''|=8\times10^{-9}$, $|f_{4}|=|f_{4}'|=|f_{4}''|=35$, $|f_{5}|=1.01\times10^{-1}$, $|f_{5}'|=5.3\times10^{-1}$, $|f_{5}''|=4\times10^{-2}$, $m_{N}=200$, $|A_{0}|=500$, $m_{E}=260$, $|m_{2}|=100$, $\chi_{m_{2}}=1.2$, $\chi_{\mu}=0.8$, $\chi_{3}=0.3$, $\chi_{3}'=0.2$, $\chi_{3}''=0.6$, $\chi_{4}=1.4$, $\chi_{4}'=1.1$, $\chi_{4}''=1.7$, $\chi_{5}=1.7$, $\chi_{5}'=0.5$, $\chi_{5}''=0.7$ and $\chi_{A_{0}}=0.4$. All masses are in GeV and phases in rad.}
\label{m0}
\end{center}
\end{figure}

In Fig.~(\ref{phase}) we investigate the effect on $\nu_3$ lifetime of the variation of  $\chi_{5}$ which is the phase of the coupling term $f_{5}$ in the neutrino mass matrix.
 The analysis is done for two values of  its magnitude $|f_5|$  (see the figure caption).
The analysis shows that the $\nu_3$ lifetime depends sensitively on the phase $\chi_5$ and also on  its magnitude.  Fig.~(\ref{phase})
 exhibits several oscillations in the lifetime as a function of $\chi_5$. \\

\begin{figure}[h]
\begin{center}
{\rotatebox{0}{\resizebox*{12cm}{!}{\includegraphics{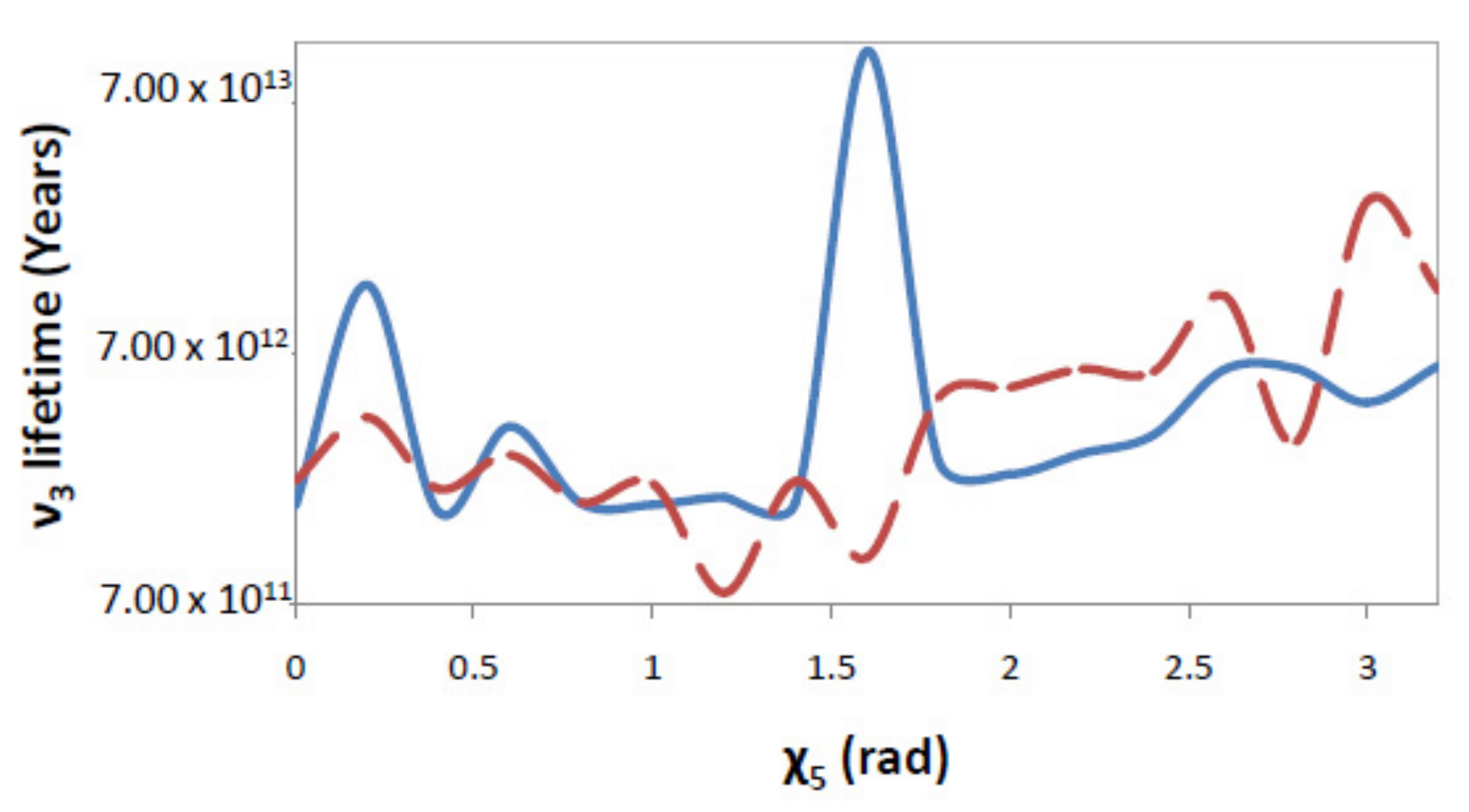}}\hglue5mm}}
\caption{Exhibition of  the dependence of  {$\nu_3$}  lifetime on the phase $\chi_{5}$ for two values of $|f_{5}|$. Solid curve is for $|f_{5}|$ = 0.1 and dashed curve is for $|f_{5}|$ = 0.05.
Other parameters have the values
$|m_{2}|=|\mu|=100$, $|f_{3}|=1.5\times10^{-7}$, $|f_{3}'|=2\times10^{-8}$, $|f_{3}''|=8\times10^{-9}$, $|f_{4}|=|f_{4}'|=|f_{4}''|=35$, $|f_{5}'|=5.3\times10^{-1}$, $|f_{5}''|=4\times10^{-2}$, $m_{N}=200$, $|A_{0}|=500$, $m_{E}=260$, $m_{0}=300$, $\tan\beta=40$, $\chi_{m_{2}}=1.2$, $\chi_{\mu}=0.8$, $\chi_{3}=0.3$, $\chi_{3}'=0.2$, $\chi_{3}''=0.6$, $\chi_{4}=1.4$, $\chi_{4}'=1.1$, $\chi_{4}''=1.7$, $\chi_{5}'=1.0$, $\chi_{5}''=0.7$ and $\chi_{A_{0}}=0.4$. All masses are in GeV and phases in rad.}
\label{phase}
\end{center}
\end{figure}

 One possible origin of such oscillations could be constructive and destructive interference
 between
 $F^{jl}_{2~left}$ and $F^{jl}_{2~right}$,  and between $F^{jl}_{3~left}$ and $F^{jl}_{3~right}$.
  Such interference was noticed and extensively studied in the context of EDMs of the quarks and the
 leptons ~\cite{Ibrahim:1997nc} (for review see ~\cite{Ibrahim:2007fb,Nath:2010zj}).
 Some numerical values are exhibited in Table~(\ref{t1}).
 Since $F_3$ is much larger than $F_2$ for this region of the parameter space, we focus on the  $F_3$ terms. Here one finds that the $F_{3 ~left}$ is larger than $F_{3~right}$ and
 further each of the terms have phases of the same sign. Thus this possibility does not appear to be the reason for large oscillations in $\nu_3$ lifetime.
 The above suggests that it is the interference in the $F_{3~left}$ terms themselves that is the origin of such rapid variation. This can come about
 because there are sixteen different contribution to $F_{3~left}$ each with their own phases and thus multiple constructive and destructive interference
 can occur which is what Fig.~(\ref{phase}) exhibits.

\begin{table}[H]
\begin{center}
\begin{tabular}{|c|c|c|}
  \hline
  $\chi_{5}$ & 0.4 rad & 1.6 rad \\
  \hline
  $F^{jl}_{2~left}$ & $(1.89\times 10^{-20})\exp(+0.34\,i)$ & $(3.56\times 10^{-21})\exp(+1.48\,i)$ \\
  $F^{jl}_{2~right}$ & $(5.53\times 10^{-21})\exp(-3.08\,i)$ & $(1.39\times 10^{-21})\exp(-1.59\,i)$ \\
  $F^{jl}_{2}(0)$ & $(1.37\times 10^{-20})\exp(+0.46\,i)$ & $(2.17\times 10^{-21})\exp(+1.73\,i)$ \\
  $F^{jl}_{3~left}$ & $(2.49\times 10^{-17})\exp(+0.63\,i)$ & $(1.59\times 10^{-18})\exp(-1.60\,i)$ \\
  $F^{jl}_{3~right}$ & $(2.68\times 10^{-18})\exp(+0.67\,i)$ & $(1.68\times 10^{-19})\exp(-1.35\,i)$ \\
  $F^{jl}_{3}(0)$ & $(2.76\times 10^{-17})\exp(+0.64\,i)$ & $(1.75\times 10^{-18})\exp(-1.58\,i)$ \\
  Decay width & $1.18\times 10^{-44}$ {GeV} & $7.58\times 10^{-48}$ {GeV} \\
  \hline
  \end{tabular}
  \caption{A list  of the right and left contributions, the form factors and the decay width of the process
  {$\nu_3\to \nu_2+ \gamma$}
 for two values of $\chi_{5}$, with $|f_{5}|$ = 0.1 GeV.}
\label{t1}
\end{center}
\end{table}

In Fig.~(\ref{A0}) we exhibit the variation of the lifetime as a function of the trilinear coupling $|A_{0}|$ for two values of $|\mu|$.  In the analysis we make the
simple approximation  $A_{\tau}=A_{E}=A_{\mu}=A_{e}=A_{0}$.\\

\begin{figure}[h]
\begin{center}
{\rotatebox{0}{\resizebox*{12cm}{!}{\includegraphics{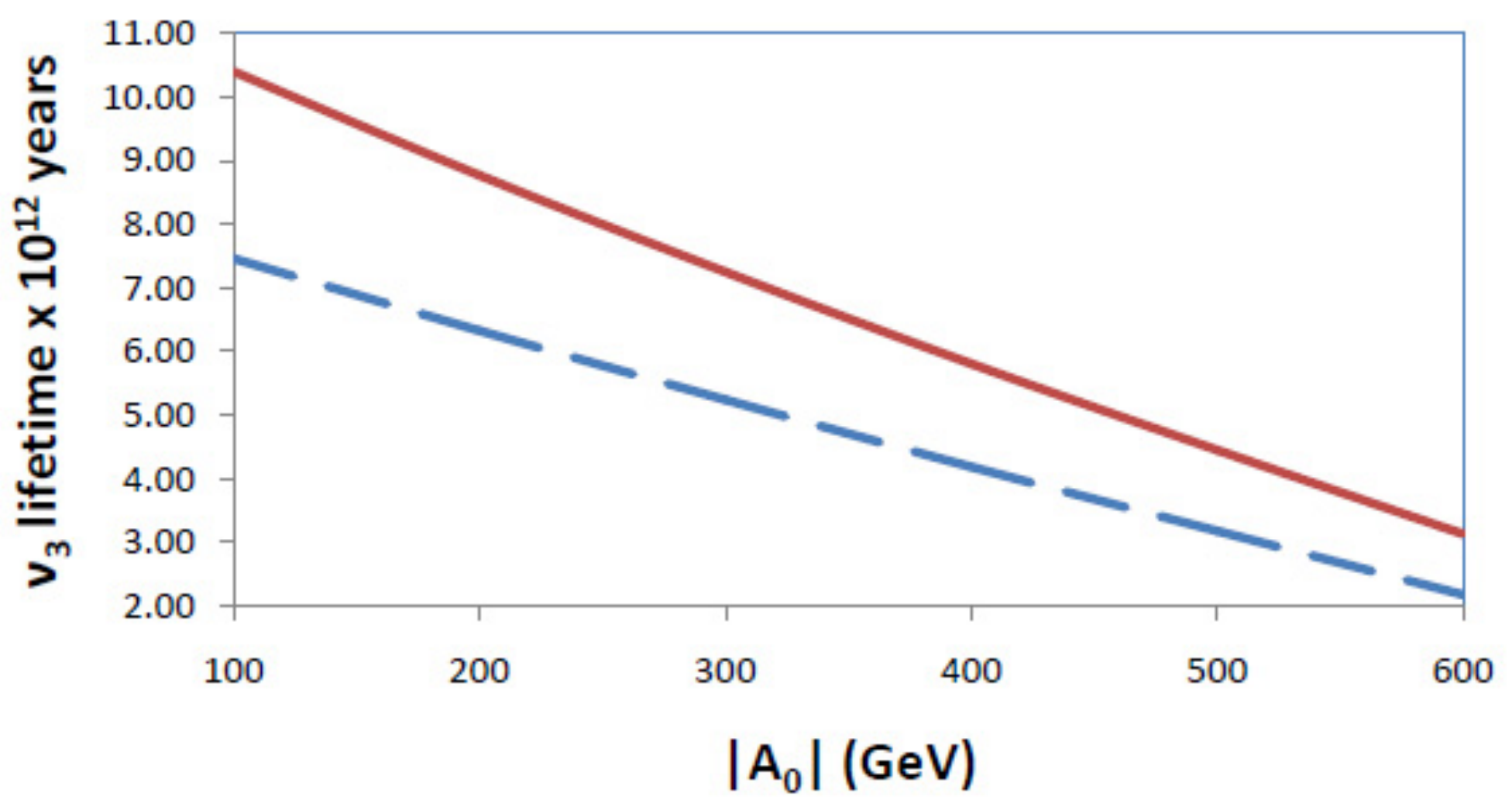}}\hglue5mm}}
\caption{Exhibition of the dependence of {$\nu_3$}
 lifetime on $|A_{0}|$ for two values of $|\mu|$. Solid curve is for $|\mu|$ = 150 and dashed curve is for $|\mu|$ = 100.  Other parameters have the values  $|m_{2}|=100$, $|f_{3}|=1.5\times10^{-7}$, $|f_{3}'|=2\times10^{-8}$, $|f_{3}''|=8\times10^{-9}$, $|f_{4}|=|f_{4}'|=|f_{4}''|=35$, $|f_{5}|=1.01\times10^{-1}$, $|f_{5}'|=5.3\times10^{-1}$, $|f_{5}''|=4\times10^{-2}$, $m_{N}=200$, $m_{E}=260$, $m_{0}=350$, $\tan\beta=50$, $\chi_{m_{2}}=1.2$, $\chi_{\mu}=0.8$, $\chi_{3}=0.3$, $\chi_{3}'=0.2$, $\chi_{3}''=0.6$, $\chi_{4}=1.4$, $\chi_{4}'=1.1$, $\chi_{4}''=1.7$, $\chi_{5}=1.7$, $\chi_{5}'=0.5$, $\chi_{5}''=0.7$ and $\chi_{A_{0}}=0.4$. All masses are in GeV and phases in rad.} \label{A0}
\end{center}
\end{figure}

Finally we discuss  the effect of $|f_{3}|$ on the tau neutrino lifetime. This analysis is exhibited in Fig.~(\ref{f3}) for two values of $\tan\beta$
(see figure caption). While $f_3$ appears both in the slepton and the neutrino mass matrix, the major effect of $f_3$ arises via the variations
in the neutrino mass matrix. In summary the analysis of Figs.(2) - (6) shows that the neutrino lifetime as low as the current experimental lower limits can be obtained in models with a vector like generation. These lifetimes are over 30 orders of magnitude smaller than in the Standard Model and thus within the reach of improved experiment.\\

\begin{figure}[H]
\begin{center}
{\rotatebox{0}{\resizebox*{12cm}{!}{\includegraphics{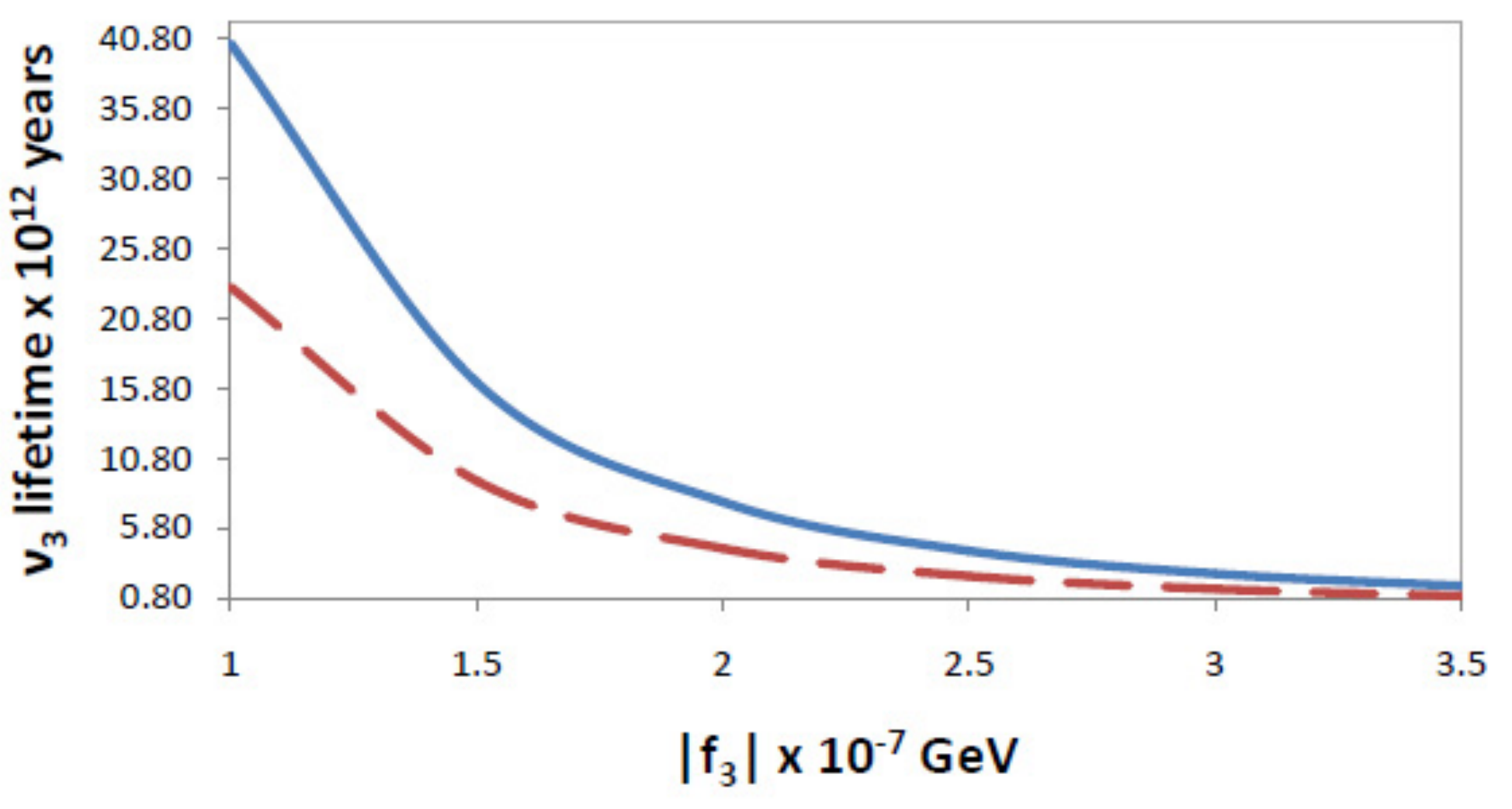}}\hglue5mm}}
\caption{Exhibition of the dependence of the {$\nu_3$} lifetime on $|f_{3}|$ for two values of $\tan\beta$.
 Solid curve is for $\tan\beta=30$ and dashed curve is for $\tan\beta=40$. Other parameters have the values
 $|m_{2}|=100$, $|\mu|=100$, $|f_{3}'|=2\times10^{-8}$, $|f_{3}''|=8\times10^{-9}$, $|f_{4}|=|f_{4}'|=|f_{4}''|=35$, $|f_{5}|=1.01\times10^{-1}$, $|f_{5}'|=5.3\times10^{-1}$, $|f_{5}''|=4\times10^{-2}$, $m_{N}=200$, $|A_{0}|=500$, $m_{E}=260$, $m_{0}=400$, $\chi_{m_{2}}=1.2$, $\chi_{\mu}=0.8$, $\chi_{3}=0.3$, $\chi_{3}'=0.2$, $\chi_{3}''=0.6$, $\chi_{4}=1.4$, $\chi_{4}'=1.1$, $\chi_{4}''=1.7$, $\chi_{5}=1.7$, $\chi_{5}'=0.5$, $\chi_{5}''=0.7$ and $\chi_{A_{0}}=0.4$. All masses are in GeV and phases in rad.}
\label{f3}
\end{center}
\end{figure}

\section{Conclusion}
Lepton flavor changing processes provide an important window to new physics beyond the Standard
Model.  In this work we have analyzed the radiative decay of the neutrinos
$\nu_i\to \nu_j \gamma$ in an  extension of the
MSSM with  a  vector like leptonic multiplet. Specifically we consider mixing between
the Standard Model generations of leptons with the mirror leptons in the vector multiplet.
It is because of these mixing which are parametrized by $f_3, f_4, f_5$,  $f_3', f_4', f_5'$,   $f_3'', f_4''$ and $ f_5''$  as
defined in Eq.(\ref{5}) that the neutrino can have a radiative decay.
The computation of the neutrino decay is done in the supersymmetric sector where we compute the
contributions to the neutrino decay  arising
 from diagrams with  exchange of charginos and staus  in the loop with the chargino or the stau
 emitting the photon.
  The effects of CP violation were also included in the analysis.   In the presence of CP phases
  both the magnetic and the electric transition dipole moments contribute to the
 neutrino lifetime. However, it is found that the electric transition dipole moment often dominates
 for moderate size CP phases in the region of the parameter space investigated.
A numerical  analysis shows that the neutrino lifetime can be smaller than the one
 predicted in the Standard Model by several orders of magnitude. Thus the Standard Model gives
 a lifetime for the decay of the heaviest neutrino $\nu_3$ so that
 $\tau_{\nu_3}^{SM}  \sim 10^{43} ~~{\rm yrs}$  for a $\nu_3$ with mass 50 meV.
 However, in the class of models  where the three generations of sleptons can
 mix with the vector like slepton generation one finds that the decay lifetime of $\nu_3$ can be
 as low as $10^{12}$ years and thus much smaller than the Standard Model prediction.
 Thus improved experiments in the future give the possibility of observation of such effects.  \\

\section{Appendix: Further details of the interactions of the vector like multiplet}
In this Appendix  we give further details of the interactions of the vector like multiplet.
The total lagrangian is constituted of ${\cal L}_F$ and  ${\cal L}_D$ where
\beq
{\cal L}_F= {\cal L}_L +{\cal L}_N.
\eeq
Here
\begin{center}
\beqn
-{\cal L}_L=\left(\frac{v^2_2 |f'_2|^2}{2} +|f_3|^2+|f_3'|^2+|f_3''|^2\right)\tilde E_R \tilde E^*_R
+\left(\frac{v^2_2 |f'_2|^2}{2} +|f_4|^2+|f_4'|^2+|f_4''|^2\right)\tilde E_L \tilde E^*_L\nonumber\\
+\left(\frac{v^2_1 |f_1|^2}{2} +|f_4|^2\right)\tilde \tau_R \tilde \tau^*_R
+\left(\frac{v^2_1 |f_1|^2}{2} +|f_3|^2\right)\tilde \tau_L \tilde \tau^*_L\nonumber\\
+\left(\frac{v^2_1 |h_1|^2}{2} +|f_4'|^2\right)\tilde \mu_R \tilde \mu^*_R
+\left(\frac{v^2_1 |h_1|^2}{2} +|f_3'|^2\right)\tilde \mu_L \tilde \mu^*_L\nonumber\\
+\left(\frac{v^2_1 |h_2|^2}{2} +|f_4''|^2\right)\tilde e_R \tilde e^*_R
+\left(\frac{v^2_1 |h_2|^2}{2} +|f_3''|^2\right)\tilde e_L \tilde e^*_L\nonumber\\
+\Bigg\{-\frac{f_1 \mu^* v_2}{\sqrt{2}} \tilde \tau_L \tilde \tau^*_R
-\frac{h_1 \mu^* v_2}{\sqrt{2}} \tilde \mu_L \tilde \mu^*_R \nonumber\\
 -\frac{f'_2 \mu^* v_1}{\sqrt{2}} \tilde E_L \tilde E^*_R
+\left(\frac{f'_2 v_2 f^*_3}{\sqrt{2}}  +\frac{f_4 v_1 f^*_1}{\sqrt{2}}\right) \tilde E_L \tilde \tau^*_L\nonumber\\
+\left(\frac{f_4 v_2 f'^*_2}{\sqrt{2}}  +\frac{f_1 v_1 f^*_3}{\sqrt{2}}\right) \tilde E_R \tilde \tau^*_R
+\left(\frac{f'_3 v_2 f'^*_2}{\sqrt{2}}  +\frac{h_1 v_1 f'^*_4}{\sqrt{2}}\right) \tilde E_L \tilde \mu^*_L
+\left(\frac{f'_2 v_2 f'^*_4}{\sqrt{2}}  +\frac{f'_3 v_1 h^*_1}{\sqrt{2}}\right) \tilde E_R \tilde \mu^*_R\nonumber\\
+\left(\frac{f''^*_3 v_2 f'_2}{\sqrt{2}}  +\frac{f''_4 v_1 h^*_2}{\sqrt{2}}\right) \tilde E_L \tilde e^*_L
+\left(\frac{f''_4 v_2 f'^*_2}{\sqrt{2}}  +\frac{f''^*_3 v_1 h^*_2}{\sqrt{2}}\right) \tilde E_R \tilde e^*_R\nonumber\\
+f'_3 f^*_3 \tilde \mu_L \tilde \tau^*_L +f_4 f'^*_4 \tilde \mu_R \tilde \tau^*_R
+f_4 f''^*_4 \tilde {e}_R \tilde{\tau}^*_R
+f''_3 f^*_3 \tilde {e}_L \tilde{\tau}^*_L\nonumber\\
+f''_3 f'^*_3 \tilde {e}_L \tilde{\mu}^*_L
+f'_4 f''^*_4 \tilde {e}_R \tilde{\mu}^*_R
-\frac{h_2 \mu^* v_2}{\sqrt{2}} \tilde{e}_L \tilde{e}^*_R
+H.c. \Bigg\}
\label{11a}
\eeqn
\end{center}
and
\beqn
-{\cal L}_N=
\left(\frac{v^2_1 |f_2|^2}{2}
 +|f_3|^2+|f_3'|^2+|f_3''|^2\right)\tilde N_R \tilde N^*_R\nonumber\\
 +\left(\frac{v^2_1 |f_2|^2}{2}+|f_5|^2+|f_5'|^2+|f_5''|^2\right)\tilde N_L \tilde N^*_L\nonumber\\
+\left(\frac{v^2_2 |f'_1|^2}{2}+|f_5|^2\right)\tilde \nu_{\tau R} \tilde \nu^*_{\tau R}
+\left(\frac{v^2_2 |f'_1|^2}{2}
+|f_3|^2\right)\tilde \nu_{\tau L} \tilde \nu^*_{\tau L}\nonumber\\
+\left(\frac{v^2_2 |h'_1|^2}{2}
+|f_3'|^2\right)\tilde \nu_{\mu L} \tilde \nu^*_{\mu L}
+\left(\frac{v^2_2 |h'_1|^2}{2}
+|f_5'|^2\right)\tilde \nu_{\mu R} \tilde \nu^*_{\mu R}\nonumber\\
+\left(\frac{v^2_2 |h'_2|^2}{2}
+|f_3''|^2\right)\tilde \nu_{e L} \tilde \nu^*_{e L}
+\left(\frac{v^2_2 |h'_2|^2}{2}
+|f_5''|^2\right)\tilde \nu_{e R} \tilde \nu^*_{e R}\nonumber\\
+\Bigg\{ -\frac{f_2 \mu^* v_2}{\sqrt{2}} \tilde N_L \tilde N^*_R
-\frac{f'_1 \mu^* v_1}{\sqrt{2}} \tilde \nu_{\tau L} \tilde \nu^*_{\tau R}
-\frac{h'_1 \mu^* v_1}{\sqrt{2}} \tilde \nu_{\mu L} \tilde \nu^*_{\mu R}\nonumber\\
+\left(\frac{f_5 v_2 f'^*_1}{\sqrt{2}}  -\frac{f_2 v_1 f^*_3}{\sqrt{2}}\right) \tilde N_L \tilde \nu^*_{\tau L}
+\left(\frac{f_5 v_1 f^*_2}{\sqrt{2}}  -\frac{f'_1 v_2 f^*_3}{\sqrt{2}}\right) \tilde N_R \tilde \nu^*_{\tau R}\nonumber\\
+\left(\frac{h'_1 v_2 f'^*_5}{\sqrt{2}}  -\frac{f'_3 v_1 f^*_2}{\sqrt{2}}\right) \tilde N_L \tilde \nu^*_{\mu L}
+\left(\frac{f''_5 v_1 f^*_2}{\sqrt{2}}  -\frac{f''^*_3 v_2 h'_2}{\sqrt{2}}\right) \tilde N_R \tilde \nu^*_{e R}\nonumber\\
+\left(\frac{h'^*_2 v_2 f''_5}{\sqrt{2}}  -\frac{f''^*_3 v_1 f_2}{\sqrt{2}}\right) \tilde N_L \tilde \nu^*_{e L}
+\left(\frac{f'_5 v_1 f^*_2}{\sqrt{2}}  -\frac{h'_1 v_2 f'^*_3}{\sqrt{2}}\right) \tilde N_R \tilde \nu^*_{\mu R}\nonumber\\
+f'_3 f^*_3 \tilde \nu_{\mu L} \tilde \nu_{\tau^*_L} +f_5 f'^*_5 \tilde \nu_{\mu R} \tilde \nu^*_{\tau R}
-\frac{h'_2 \mu^* v_1}{\sqrt{2}} \tilde{\nu}_{e L} \tilde{\nu}^*_{e R}\nonumber\\
+f''_3 f^*_3   \tilde{\nu}_{e L} \tilde{\nu}^*_{\tau L}
+f_5 f''^*_5   \tilde{\nu}_{e R} \tilde{\nu}^*_{\tau R}
+f''_3 f'^*_3   \tilde{\nu}_{e L} \tilde{\nu}^*_{\mu L}
+f'_5 f''^*_5   \tilde{\nu}_{e R} \tilde{\nu}^*_{\mu R}
+H.c. \Bigg\}.
\label{11b}
\eeqn

Similarly the mass terms arising from the D term are given by
\beqn
-{\cal L}_D=\frac{1}{2} m^2_Z \cos^2\theta_W \cos 2\beta \{\tilde \nu_{\tau L} \tilde \nu^*_{\tau L} -\tilde \tau_L \tilde \tau^*_L
+\tilde \nu_{\mu L} \tilde \nu^*_{\mu L} -\tilde \mu_L \tilde \mu^*_L\nonumber\\
+\tilde \nu_{e L} \tilde \nu^*_{e L} -\tilde e_L \tilde e^*_L
+\tilde E_R \tilde E^*_R -\tilde N_R \tilde N^*_R\}\nonumber\\
+\frac{1}{2} m^2_Z \sin^2\theta_W \cos 2\beta \{\tilde \nu_{\tau L} \tilde \nu^*_{\tau L}\nonumber\\
 +\tilde \tau_L \tilde \tau^*_L
+\tilde \nu_{\mu L} \tilde \nu^*_{\mu L} +\tilde \mu_L \tilde \mu^*_L
+\tilde \nu_{e L} \tilde \nu^*_{e L} +\tilde e_L \tilde e^*_L
\nonumber\\
-\tilde E_R \tilde E^*_R -\tilde N_R \tilde N^*_R +2 \tilde E_L \tilde E^*_L -2 \tilde \tau_R \tilde \tau^*_R
-2 \tilde \mu_R \tilde \mu^*_R -2 \tilde e_R \tilde e^*_R
\}.
\label{12}
\eeqn
In addition we have the following set of soft breaking terms
\beqn
V_{soft}=\tilde M^2_{\tau L} \tilde \psi^{i*}_{\tau L} \tilde \psi^i_{\tau L}
+\tilde M^2_{\chi} \tilde \chi^{ci*} \tilde \chi^{ci}\nonumber\\
+\tilde M^2_{\mu L} \tilde \psi^{i*}_{\mu L} \tilde \psi^i_{\mu L}
+\tilde M^2_{e L} \tilde \psi^{i*}_{e L} \tilde \psi^i_{e L}
+\tilde M^2_{\nu_\tau} \tilde \nu^{c*}_{\tau L} \tilde \nu^c_{\tau L}
\nonumber\\
 +\tilde M^2_{\nu_\mu} \tilde \nu^{c*}_{\mu L} \tilde \nu^c_{\mu L}
+\tilde M^2_{\nu_e} \tilde \nu^{c*}_{e L} \tilde \nu^c_{e L}
+\tilde M^2_{\tau} \tilde \tau^{c*}_L \tilde \tau^c_L \nonumber\\
+\tilde M^2_{\mu} \tilde \mu^{c*}_L \tilde \mu^c_L
+\tilde M^2_{e} \tilde e^{c*}_L \tilde e^c_L
+\tilde M^2_E \tilde E^*_L \tilde E_L
 + \tilde M^2_N \tilde N^*_L \tilde N_L \nonumber\\
+\epsilon_{ij} \{f_1 A_{\tau} H^i_1 \tilde \psi^j_{\tau L} \tilde \tau^c_L
-f'_1 A_{\nu_\tau} H^i_2 \tilde \psi ^j_{\tau L} \tilde \nu^c_{\tau L}\nonumber\\
+h_1 A_{\mu} H^i_1 \tilde \psi^j_{\mu L} \tilde \mu^c_L
-h'_1 A_{\nu_\mu} H^i_2 \tilde \psi ^j_{\mu L} \tilde \nu^c_{\mu L}\nonumber\\
+h_2 A_{e} H^i_1 \tilde \psi^j_{e L} \tilde e^c_L
-h'_2 A_{\nu_e} H^i_2 \tilde \psi ^j_{e L} \tilde \nu^c_{e L}
\nonumber\\
+f_2 A_N H^i_1 \tilde \chi^{cj} \tilde N_L
-f'_2 A_E H^i_2 \tilde \chi^{cj} \tilde E_L +H.c.\}
\label{13}
\eeqn
From ${\cal L}_{F,D}$ and
by giving the neutral Higgs their vacuum expectation values in  $V_{soft}$ we can produce
the mass squared   matrix $M^2_{\tilde \tau}$  in the basis $(\tilde  \tau_L, \tilde E_L, \tilde \tau_R,
\tilde E_R, \tilde \mu_L, \tilde \mu_R, \tilde e_L, \tilde e_R)$. We  label the matrix  elements of these as $(M^2_{\tilde \tau})_{ij}= M^2_{ij}$ where
\beqn
M^2_{11}=\tilde M^2_{\tau L} +\frac{v^2_1|f_1|^2}{2} +|f_3|^2 -m^2_Z \cos 2 \beta \left(\frac{1}{2}-\sin^2\theta_W\right), \nonumber\\
M^2_{22}=\tilde M^2_E +\frac{v^2_2|f'_2|^2}{2}+|f_4|^2 +|f'_4|^2+|f''_4|^2 +m^2_Z \cos 2 \beta \sin^2\theta_W, \nonumber\\
M^2_{33}=\tilde M^2_{\tau} +\frac{v^2_1|f_1|^2}{2} +|f_4|^2 -m^2_Z \cos 2 \beta \sin^2\theta_W, \nonumber\\
M^2_{44}=\tilde M^2_{\chi} +\frac{v^2_2|f'_2|^2}{2} +|f_3|^2 +|f'_3|^2+|f''_3|^2 +m^2_Z \cos 2 \beta \left(\frac{1}{2}-\sin^2\theta_W\right), \nonumber\\
M^2_{55}=\tilde M^2_{\mu L} +\frac{v^2_1|h_1|^2}{2} +|f'_3|^2 -m^2_Z \cos 2 \beta \left(\frac{1}{2}-\sin^2\theta_W\right), \nonumber\\
M^2_{66}=\tilde M^2_{\mu} +\frac{v^2_1|h_1|^2}{2}+|f'_4|^2 -m^2_Z \cos 2 \beta \sin^2\theta_W, \nonumber\\
M^2_{77}=\tilde M^2_{e L} +\frac{v^2_1|h_2|^2}{2}+|f''_3|^2 -m^2_Z \cos 2 \beta \left(\frac{1}{2}-\sin^2\theta_W\right), \nonumber\\
M^2_{88}=\tilde M^2_{e} +\frac{v^2_1|h_2|^2}{2}+|f''_4|^2 -m^2_Z \cos 2 \beta \sin^2\theta_W, \nonumber\\
M^2_{12}=M^{2*}_{21}=\frac{ v_2 f'_2f^*_3}{\sqrt{2}} +\frac{ v_1 f_4 f^*_1}{\sqrt{2}} ,\nonumber\\
M^2_{13}=M^{2*}_{31}=\frac{f^*_1}{\sqrt{2}}(v_1 A^*_{\tau} -\mu v_2),\nonumber\\
M^2_{14}=M^{2*}_{41}=0, M^2_{15} =M^{2*}_{51}=f'_3 f^*_3,\nonumber\\
 M^{2*}_{16}= M^{2*}_{61}=0,  M^{2*}_{17}= M^{2*}_{71}=f''_3 f^*_3,  M^{2*}_{18}= M^{2*}_{81}=0,
M^2_{23}=M^{2*}_{32}=0,\nonumber\\
M^2_{24}=M^{2*}_{42}=\frac{f'^*_2}{\sqrt{2}}(v_2 A^*_{E} -\mu v_1),  M^2_{25} = M^{2*}_{52}= \frac{ v_2 f'_3f'^*_2}{\sqrt{2}} +\frac{ v_1 h_1 f^*_4}{\sqrt{2}} ,\nonumber\\
 M^2_{26} =M^{2*}_{62}=0,  M^2_{27} =M^{2*}_{72}=  \frac{ v_2 f''_3f'^*_2}{\sqrt{2}} +\frac{ v_1 h_1 f'^*_4}{\sqrt{2}},  M^2_{28} =M^{2*}_{82}=0, \nonumber\\
M^2_{34}=M^{2*}_{43}= \frac{ v_2 f_4 f'^*_2}{\sqrt{2}} +\frac{ v_1 f_1 f^*_3}{\sqrt{2}}, M^2_{35} =M^{2*}_{53} =0, M^2_{36} =M^{2*}_{63}=f_4 f'^*_4,\nonumber\\
 M^2_{37} =M^{2*}_{73} =0,  M^2_{38} =M^{2*}_{83} =f_4 f''^*_4,
M^2_{45}=M^{2*}_{54}=0, M^2_{46}=M^{2*}_{64}=\frac{ v_2 f'_2 f'^*_4}{\sqrt{2}} +\frac{ v_1 f'_3 h^*_1}{\sqrt{2}}, \nonumber\\
 M^2_{47} =M^{2*}_{74}=0,  M^2_{48} =M^{2*}_{84}=  \frac{ v_2 f'_2f''^*_4}{\sqrt{2}} +\frac{ v_1 f''_3 h^*_2}{\sqrt{2}},\nonumber\\
M^2_{56}=M^{2*}_{65}=\frac{h^*_1}{\sqrt{2}}(v_1 A^*_{\mu} -\mu v_2),
 M^2_{57} =M^{2*}_{75}=f''_3 f'^*_3,  M^2_{58} =M^{2*}_{85}=0,  M^2_{67} =M^{2*}_{76}=0,\nonumber\\
 M^2_{68} =M^{2*}_{86}=f'_4 f''^*_4,  M^2_{78}=M^{2*}_{87}=\frac{h^*_2}{\sqrt{2}}(v_1 A^*_{e} -\mu v_2)
\label{14}
\eeqn

Here the terms $M^2_{11}, M^2_{13}, M^2_{31}, M^2_{33}$ arise from soft
breaking in the  sector $\tilde \tau_L, \tilde \tau_R$,
the terms $M^2_{55}, M^2_{56}, M^2_{65}, M^2_{66}$ arise from soft
breaking in the  sector $\tilde \mu_L, \tilde \mu_R$,
the terms $M^2_{77}, M^2_{78}, M^2_{87}, M^2_{88}$ arise from soft
breaking in the  sector $\tilde e_L, \tilde e_R$
 and
the terms
$M^2_{22}, M^2_{24},$  $M^2_{42}, M^2_{44}$ arise from soft
breaking in the  sector $\tilde E_L, \tilde E_R$. The other terms arise  from mixing between the staus, smuons and
the mirrors.  We assume that all the masses are of the electroweak size
so all the terms enter in the mass squared  matrix.  We diagonalize this hermitian mass squared  matrix  by the
 unitary transformation
$
 \tilde D^{\tau \dagger} M^2_{\tilde \tau} \tilde D^{\tau} = diag (M^2_{\tilde \tau_1},
M^2_{\tilde \tau_2}, M^2_{\tilde \tau_3},  M^2_{\tilde \tau_4},  M^2_{\tilde \tau_5},  M^2_{\tilde \tau_6},  M^2_{\tilde \tau_7},  M^2_{\tilde \tau_8} )$. {For a further clarification of the notation see~\cite{Ibrahim:2012ds})}.\\

\noindent
{\em Acknowledgments}:
One of the authors (PN) acknowledges the hospitality of KITP, Santa Barbara, where part of this
work was done. This research was supported in part by the National Science Foundation under Grant Nos.
PHY-0757959, PHY-0704067 and NSF PHY11-25915.

\end{document}